\documentclass[conference, pdftex]{IEEEtran}
\IEEEoverridecommandlockouts

\usepackage{mgmmathtool}
\usepackage{color}
\usepackage{siunitx}

\newcommand{\Hline}{\noalign{\hrule height 0.4mm}}
\usepmat

\usepackage{cite}
\usepackage{amsmath,amssymb,amsfonts}
\usepackage{algorithmic}
\usepackage{graphicx}
\usepackage{textcomp}
\def\BibTeX{{\rm B\kern-.05em{\sc i\kern-.025em b}\kern-.08em
    T\kern-.1667em\lower.7ex\hbox{E}\kern-.125emX}}
\begin{document}
\setlength{\abovedisplayskip}{3.5pt} 
\setlength{\belowdisplayskip}{3.5pt} 

%
\title{%
Independent Deeply Learned Matrix Analysis
for Multichannel Audio Source Separation
}

\author{%
\IEEEauthorblockN{Shinichi Mogami,
Hayato Sumino,
Daichi Kitamura,\\
Norihiro Takamune,
Shinnosuke Takamichi,
Hiroshi Saruwatari
}
\IEEEauthorblockA{\textit{The University of Tokyo}}
\and
\IEEEauthorblockN{Nobutaka Ono}
\IEEEauthorblockA{\textit{Tokyo Metropolitan University}}
}

\maketitle

\begin{abstract}
In this paper, we address a multichannel audio source separation task and propose a new efficient method called independent deeply learned matrix analysis (IDLMA).
IDLMA estimates the demixing matrix in a blind manner and updates the time-frequency structures of each source using a pretrained deep neural network (DNN).
Also, we introduce a complex Student's $t$-distribution as a generalized source generative model including both complex Gaussian and Cauchy distributions.
Experiments are conducted using music signals with a training dataset, and the
results show the validity of the proposed method in terms of separation accuracy and computational cost.
\end{abstract}



\section{Introduction}

Blind source separation (BSS) is a technique for
extracting specific sources from an observed multichannel mixture signal
without knowing a priori information about the mixing system.
The most commonly used algorithm for BSS
in the (over)determined case ($\text{number of microphones}\geq\text{number of sources}$)
is independent component analysis~(ICA)~\cite{PComon1994_ICA}.
Recently,
\emph{independent low-rank matrix analysis~(ILRMA)}%
~\cite{DKitamura2016_ILRMA,Kitamura2018_ILRMAbook},
which is a unification of
independent vector analysis~(IVA)~\cite{TKim2007_IVA}
and nonnegative matrix factorization~(NMF)~\cite{DDLee1999_NMF},
was proposed as a state-of-the-art BSS method.
ILRMA assumes both statistical independence between sources and a low-rank time-frequency structure for each source,
and the frequency-wise demixing matrices are estimated
without encountering the permutation problem.
The source generative model assumed in ILRMA was generalized
from a complex Gaussian distribution~\cite{DKitamura2016_ILRMA}
to complex Student's $t$-distribution ($t$-ILRMA)~\cite{Mogami2017_tILRMA}
for more robust BSS.
As a more general framework, in~\cite{Lopez2015_DesignIVA},
demixing matrix optimization based on
a given spectrogram estimate for the source
was proposed,
showing that the precise source spectrogram model
enables accurate spatial model estimation.

In the underdetermined case ($\text{number of microphones}<\text{number of sources}$),
the Duong model~\cite{Duong2010_SpatialMatrix} is a commonly used framework.
In the Duong model, frequency-wise spatial covariances,
which encode source locations and their spatial spreads,
are estimated by an expectation-maximization (EM) algorithm,
where the permutation problem must be solved after the optimization.
Similarly to ILRMA, an NMF-based low-rank assumption is employed in the Duong model
to automatically solve the permutation problem,
resulting in multichannel NMF (MNMF)%
~\cite{Ozerov2012_MNMF,HSawada2013_MNMF}.
Note that these algorithms formulate a \emph{mixing} model,
whereas ICA-based methods including ILRMA estimate a \emph{demixing} model
for the separation by focusing only on the determined case.
It has been experimentally confirmed that the optimization of a demixing model
is more efficient and numerically stable than that of a mixing model~\cite{DKitamura2016_ILRMA}.

In supervised (informed) source separation,
deep neural network (DNN) has shown promising performance
in both single-channel~\cite{Grais2013_SingleSSwDNN}
and multichannel source separation~\cite{Araki2015_MultiSSwDNN}.
In fact, when sufficient data of the audio sources are available,
DNN can effectively model their time-frequency structures.
However, it is almost impossible
to compose an appropriate and generalized spatial model with DNN
from training data observed in a multichannel format.
This is because the spatial model depends on many factors,
including source and microphone locations, the recording room, and reverberation.
Therefore, it is reasonable
to combine a pretrained DNN source model and a blind estimation of the spatial model.
Nugraha et al. proposed a DNN-based multichannel source separation framework~\cite{Arie2016_DuongDNN}
using the Duong model (hereafter referred to as \emph{Duong+DNN}).
Although this is a convincing approach,
a large computational cost is required to estimate the spatial covariance (the EM algorithm in the Duong model)
and the performance is not satisfactory
owing to the difficulty of parameter optimization.

In this paper, we unify
the ICA-based blind estimation of the demixing matrix
and the DNN-based supervised update of the source spectrogram model.
In the proposed method,
we introduce a complex Student's $t$-distribution
as a generalized source generative model,
and the demixing matrix (spatial model) is
efficiently optimized using a majorization-minimization (MM) algorithm~\cite{DRHunter2000_MMalgorithm}.
Since the proposed method utilizes a time-frequency spectrogram matrix
estimated by DNN to optimize the spatial model,
we call this method \emph{independent deeply learned matrix analysis (IDLMA)}.
Table~\ref{tab:overview-ass} shows the relationship between the existing and proposed methods.
The spatial model is blindly estimated in all the methods,
while the source spectrogram model is estimated by DNN in Duong+DNN and the proposed IDLMA.

\section{Conventional Method}
\subsection{Formulation}
Let $N$ and $M$ be the numbers of sources and channels, respectively.
The short-time Fourier transform (STFT) of the multichannel source,
observed, and estimated signals are defined as
$
\bm s_{ij}=(s_{ij1}, \ldots, s_{ijN})\T,
\bm x_{ij}=(x_{ij1}, \ldots, x_{ijM})\T,
$
and
$
\bm y_{ij}=(y_{ij1}, \ldots, y_{ijN})\T
$,
where
$i=1,\ldots, I;
 j=1,\ldots, J;
 n=1,\ldots, N;$ and
$m=1,\ldots, M$
are the integral indexes of the frequency bins, time frames, sources, and channels, respectively
,
and ${}\T$ denotes the transpose.
We also denote these spectrograms
as $
\bm S_n\in\cset^{I\times J},
\bm X_m\in\cset^{I\times J}
$,
and
$
\bm Y_n\in\cset^{I\times J}
$,
whose elements are $s_{ijn}, x_{ijn}$, and $y_{ijn}$, respectively.
In ILRMA,
the following mixing system is assumed:
\begin{align}
\bm x_{ij} = \bm A_i \bm s_{ij},
\label{eq:mixing-system}
\end{align}
where $\bm A_i = (\bm a_{i1}, \ldots, \bm a_{iN})\in \cset^{M\times N}$
is a frequency-wise mixing matrix
and $\bm a_{in}$ is the steering vector for the $n$th source.
The assumption of the mixing system~\eqref{eq:mixing-system}
corresponds to restricting the spatial covariance in the Duong model to a rank-1 matrix~\cite{Duong2010_SpatialMatrix}.
When $M=N$ and $\bm{A}_i$ is not a singular matrix, the estimated signal $\bm{y}_{ij}$ can be represented as
\begin{align}
\bm y_{ij} = \bm W_i \bm x_{ij},
\end{align}
where $\bm{W}_i= \bm A_{i}^{-1} = (\bm w_{i1},\ldots, \bm w_{iN})\Ht$ is the demixing matrix,
$\bm{w}_{in}$ is the demixing filter for the $n$th source,
and ${}\Ht$ denotes the Hermitian transpose.
ILRMA estimates both $\bm W_i$ and $\bm y_{ij}$ from only the observation $\bm x_{ij}$
assuming statistical independence between $s_{ijn}$ and $s_{ijn'}$, where $n\ne n'$.

\begin{table}[tp]
\centering
\caption{%
Classification of multichannel source separation methods
\label{tab:overview-ass}%
}
{
\begin{tabular}{c|cc}
\Hline
&\multicolumn{2}{c}{Source spectrogram model}\\\cline{2-3}
& Blind & Supervised\\\hline
Mixing model & MNMF~\cite{AOzerov2010_MNMF,HSawada2013_MNMF}
& Duong+DNN~\cite{Arie2016_DuongDNN} \\
Demixing model & ILRMA~\cite{DKitamura2016_ILRMA,Mogami2017_tILRMA}
& \textbf{Proposed IDLMA} \\
\Hline
\end{tabular}}
\end{table}

\subsection{ILRMA and Its Generalization with Student's $t$-distribution}
In \cite{DKitamura2016_ILRMA,Kitamura2018_ILRMAbook},
the following time-frequency-varying complex Gaussian source generative model is assumed (hereafter referred to as Gauss-ILRMA):
\begin{align}
\prod_{i,j} p(y_{ijn})
&= \prod_{i,j} \dfrac1{\pi {\sigma_{ijn}}^2}
\exp\p[4]{ -\dfrac{\abs{y_{ijn}}^2}{{\sigma_{ijn}}^2} },
\label{eq:is-ILRMA-generative-model}\\
{\sigma_{ijn}}^2 &= \sum_k t_{ikn}v_{kjn},
\label{eq:is-ILRMA-source-model}
\end{align}
where $\sigma_{ijn}$ is the variance (source spectrogram model),
$k=1,\ldots,K$ is the index of the bases,
and $t_{ikn}$ and $v_{kjn}$ are the parameters in the NMF-based low-rank model.
We also denote the variance matrix as $\bm\Sigma_{n}\in\rset_{\ge0}^{I\times J}$,
whose elements are $\sigma_{ijn}$.
In $t$-ILRMA~\cite{Mogami2017_tILRMA}, \eqref{eq:is-ILRMA-generative-model}
is generalized to a complex Student's $t$-distribution as follows:
\begin{align}
\prod_{i,j} p(y_{ijn})
&= \prod_{i,j} \dfrac1{\pi {\sigma_{ijn}}^2}
\p[4]{
  1 + \dfrac2\nu \dfrac{\abs{y_{ijn}}^2}{{\sigma_{ijn}}^2}
}^{-\frac{2+\nu}{2}},
\label{eq:t-ILRMA-generative-model}
\\
{\sigma_{ijn}}^p &= \sum_k t_{ikn}v_{kjn},
\label{eq:t-ILRMA-source-model}
\end{align}
where $\nu$ is the degree-of-freedom parameter
and $p$ is the domain parameter.
When $\nu\to\infty$ and $p=2$,
\eqref{eq:t-ILRMA-generative-model} and \eqref{eq:t-ILRMA-source-model}
become~\eqref{eq:is-ILRMA-generative-model} and \eqref{eq:is-ILRMA-source-model}, respectively.
Also, \eqref{eq:t-ILRMA-generative-model} with $\nu=1$ represents the Cauchy-distribution likelihood.
The demixing matrix $\bm{W}_i$ and NMF source model $t_{ikn}v_{kjn}$
can be optimized in the maximum-likelihood (ML) sense
on the basis of \eqref{eq:is-ILRMA-generative-model} or \eqref{eq:t-ILRMA-generative-model}.
Since the low-rank structure of
{\small $\abs*{\bm{Y}_n}^{\lower1.0pt\hbox{$\scriptstyle.2$}}$}
is ensured by the NMF source model, the permutation problem can be avoided,
where $\abs{\cdot}^{.p}$ 
for matrices denotes the element-wise absolute and $p$th-power operations.


\section{Proposed Method}
\subsection{Motivation}
The NMF source model in ILRMA is effective for some sources
that have a low-rank time-frequency structure.
However, this source spectrogram model is not always valid.
For example, speech signals have continuously varying spectra,
which cannot be efficiently modeled by NMF, and the separation performance of ILRMA is degraded for such sources.
If sufficient training data for each source can be prepared in advance,
it is possible to construct a suitable source spectrogram model
by employing DNN~\cite{Grais2013_SingleSSwDNN}.
On the other hand, since the spatial parameters depend on many factors,
it is simply impractical to train a general spatial model with DNN
even if huge amounts of multichannel observation data are available;
therefore, the spatial parameters should be estimated \textit{blindly}.

In this paper, we propose a new framework, IDLMA,
which combines
the ICA-based blind estimation of demixing matrix $\bm W_i$
and the supervised learning of variance matrix $\bm\Sigma_{n}$ based on DNN,
where the loss function in DNN is designed to maximize the likelihood of the source generative model.
In addition, similarly to $t$-ILRMA,
we use a generalized model based on a complex Student's $t$-distribution including both Gaussian and Cauchy distributions.
Duong+DNN also employs DNN that maximizes the likelihood of the Gaussian or Cauchy distribution.
However, since the mixing model (spatial covariance) in Duong+DNN is defined by only the Gaussian model,
the estimations of the spectral and spatial parameters are inconsistent.
In the proposed method,
this conflict is resolved by modeling the spatial parameters
with the Student's $t$-distribution model
and deriving its optimization algorithm fully consistently in the ML sense.

\begin{figure*}[tp]
\vspace{1ex}
\centering
\includegraphics[width=.9\textwidth]{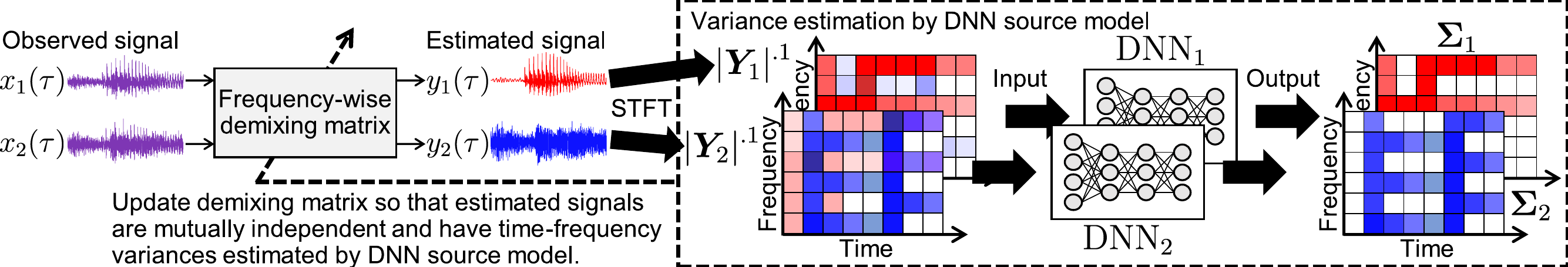}
\caption{Principle of source separation based on
IDLMA in case of $N = M = 2$.
\label{fig:principle-idlma}%
}
\vspace{-3ex}
\end{figure*}

\subsection{Cost Function in IDLMA}
Let $\mathrm{DNN}_n$ be the DNN source model
that enhances the $n$th source component from a mixture signal,
namely, the variance matrix $\bm\Sigma_n$ is estimated by $\mathrm{DNN}_n$,
and these DNN source models are trained in advance.
Fig.~\ref{fig:principle-idlma} shows the principle of the separation mechanism in the proposed IDLMA.

On the basis of~\eqref{eq:is-ILRMA-generative-model},
the cost function
(negative log-likelihood of $\bm x_{ij} = \bm W_i^{-1} \bm y_{ij}$)
in IDLMA with the complex Gaussian distribution (Gauss-IDLMA) is obtained as
\begin{align}
\mathcal{L}_{\mathrm{Gauss}}
=\sum_{i,j,n}\lr[]{
\dfrac{
  \abs{y_{ijn}}^2
  }{
  {\sigma_{ijn}}^2
}
+ 2\log \sigma_{ijn}
}
- 2J\sum_i \log \abs{\det \bm W_i}
,
\label{eq:is-idlma-cost-function}
\end{align}
and \eqref{eq:is-idlma-cost-function} can be generalized with \eqref{eq:t-ILRMA-generative-model} ($t$-IDLMA) as
\begin{align}
\mathcal{L}_t &=\sum_{i,j,n}\lr[]{
\p{1 + \dfrac \nu 2} \log\p[4]{
1 + \dfrac 2 \nu \dfrac{
  \abs{y_{ijn}}^2
  }{
  {\sigma_{ijn}}^2
}}
+ 2\log \sigma_{ijn}
}\notag\\
&\phantom{{}={}} - 2J\sum_i \log \abs{\det \bm W_i}
,
\label{eq:t-idlma-cost-function}
\end{align}
where $y_{ijn}=\bm w_{in}\Ht \bm x_{ij}$.
Note that $\mathcal{L}_t$ converges to $\mathcal{L}_{\mathrm{Gauss}}$ when $\nu\to\infty$.

\subsection{Update Rule of Source Spectrogram Model Based on DNN}
$\mathrm{DNN}_n$ is trained so that the source spectrogram $\abs*{\tilde{\bm{S}}_n}^{.1}$ is predicted
from an input mixture spectrogram $|\tilde{\bm{X}}|^{.1}$,
where $\tilde{\bm{S}}_n \in \cset^{I\times J}$ and $\tilde{\bm{X}} \in\cset^{I\times J}$
are source and mixture spectrograms in the training data, respectively.
When we define the output spectrogram as $\bm D_n = \mathrm{DNN}_n(\abs*{\tilde{\bm X}}^{.1})\in\rset_{\ge0}^{I\times J}$,
the loss function of $\mathrm{DNN}_n$ for Gauss-IDLMA is defined as
\begin{align}
\mathrm{L}_{\mathrm{Gauss}}(\bm D_n)
=\sum_{i,j}\p{
\dfrac{
  \abs{\tilde{s}_{ijn}}^2 + \delta_1
  }{
  {d_{ijn}}^2 + \delta_1
} - \log \dfrac{
  \abs{\tilde{s}_{ijn}}^2 + \delta_1
  }{
  {d_{ijn}}^2 + \delta_1
} - 1
},
\label{eq:is-idlma-dnn-loss}
\end{align}
where $\tilde s_{ijn}$ and $d_{ijn}$ are the elements of $\tilde{\bm S}_n$ and ${\bm D}_n$, respectively,
and $\delta_1$ is a small value to avoid division by zero~\cite{Arie2016_DuongDNN}.
Also, the loss function of $\mathrm{DNN}_n$ for $t$-IDLMA is defined as
\begin{align}
\mathrm{L}_t(\bm D_{n}) &= \sum_{i,j}\Biggl[
\p{1 + \dfrac \nu 2} \log\p[4]{
1 + \dfrac 2 \nu \dfrac{
  \abs{\tilde s_{ijn}}^2 + \delta_1
  }{
  {d_{ijn}}^2 + \delta_1
}} \notag\\
&\phantom{{}=\sum_{i,j}[}+ \log ({d_{ijn}}^2 + \delta_1)
\Biggr].
\label{eq:t-idlma-dnn-loss}
\end{align}
Since minimizing \eqref{eq:is-idlma-dnn-loss} or \eqref{eq:t-idlma-dnn-loss}
is equivalent to the ML estimation of
$\sigma_{ijn}$ in \eqref{eq:is-idlma-cost-function} or \eqref{eq:t-idlma-cost-function},
$\mathrm{DNN}_n$ can be interpreted as the proper source generative model based on
\eqref{eq:is-ILRMA-generative-model} or \eqref{eq:t-ILRMA-generative-model}, respectively.
Similarly to \eqref{eq:t-idlma-cost-function}, $\mathrm{L}_t(\bm D_{n})$
converges to $\mathrm{L}_{\mathrm{Gauss}}(\bm D_{n})$ up to a constant when $\nu\to\infty$.

The variance matrix
is updated by the trained $\mathrm{DNN}_n$ as
\begin{align}
\abs*{{\bm \Sigma}_n}^{.1} &\gets \mathrm{DNN}_n (\abs*{{\bm Y}_n}^{.1}),
\label{eq:update-dnn-1}
\\
\sigma_{ijn}&\gets \max( \sigma_{ijn}, \3),
\label{eq:update-dnn-2}
\end{align}
where
$\3$ is a small value to increase the numerical stability of the spatial update described in Sect.~\ref{sse:update-w}.
The DNN architectures used in this paper are described in detail in Sect.~\ref{sse:dnn-spectral-models}.

\subsection{Update Rule of Demixing Matrix\label{sse:update-w}}
The demixing matrix $\bm W_i$ can be optimized
while taking the statistical independence between sources
and the variance matrix $\bm\Sigma_n$ into account
on the basis of \eqref{eq:is-ILRMA-generative-model} or \eqref{eq:t-ILRMA-generative-model}.
In Gauss-IDLMA,
$\bm W_i$ can be updated by applying
iterative projection (IP)~\cite{NOno2011_AuxIVA} to \eqref{eq:is-idlma-cost-function},
where IP is a fast and stable optimization algorithm 
that can be applied to the sum of $\abs{\bm w_{in}\Ht \bm x_{ij}}^2$ and $-\log\abs{\det \bm W_i}$.
In $t$-IDLMA, IP cannot be applied to \eqref{eq:t-idlma-cost-function}
because $\abs{\bm w_{in}\Ht \bm x_{ij}}^2$ is intrinsic in the logarithm function. 
Therefore, we apply an MM algorithm~\cite{DRHunter2000_MMalgorithm} to derive the update rule of $\bm w_{in}$.

To design a majorization function for \eqref{eq:t-idlma-cost-function}, we apply the tangent line inequality
\begin{align}
\log z \le \dfrac1{\alpha}(z - \alpha) + \log \alpha
\label{eq:tangent-line-inequality}
\end{align}
to the logarithm term in \eqref{eq:t-idlma-cost-function}, where $z>0$ is the original variable and $\alpha>0$ is an auxiliary variable.
The majorization function can be designed as
\begin{align}
\mathcal{L}_t&\le \sum_{i,j,n}
\Biggl[
\p{1 + \dfrac \nu 2} \dfrac1{\alpha_{ijn}}
\p[4]{
1 + \dfrac 2 \nu
\dfrac{
  \abs{y_{ijn}}^2
  }{
  {\sigma_{ijn}}^2
}
- \alpha_{ijn}
}\notag\\
&\phantom{{}=\sum[} + \p{1 + \dfrac \nu 2} \log \alpha_{ijn}
+ 2\log \sigma_{ijn}
\Biggr]\notag\\
&\phantom{{}={}} - 2J\sum_i \log \abs{\det \bm W_i} \notag\\
&=: \mathcal{L}_t^+,
\label{eq:w-update-majorization}
\end{align}
where $\alpha_{ijn}$ is the auxiliary variable, and $\mathcal{L}_t$ and
$\mathcal{L}_t^+$ become equal only when
\begin{align}
\alpha_{ijn} = 1 + \dfrac2\nu \dfrac{\abs{y_{ijn}}^2}{{\sigma_{ijn}}^2}.
\label{eq:alpha-equal-condition}
\end{align}
We can apply IP in analogy with the derivation in Gauss-ILRMA.
The majorization function~\eqref{eq:w-update-majorization} is reformulated as
\begin{align}
\mathcal{L}_t^+&=
J\sum_{i,n}\bm w_{in}\Ht \bm U_{in} \bm w_{in}
- 2J\sum_i \log \abs{\det \bm W_i} + \mathrm{const.}, \\
\bm U_{in} &= \dfrac1J \p{1 + \dfrac2\nu}
\sum_{j} \dfrac1{\alpha_{ijn}{\sigma_{ijn}}^2} \bm x_{ij} \bm x_{ij}\Ht.
\end{align}
By applying IP and substituting~\eqref{eq:alpha-equal-condition},
the demixing filter $\bm w_{in}$ can be updated as follows:
\begin{align}
\bm w_{in} &\gets (\bm W_i\bm U_{in})^{-1}\bm e_n,
\label{eq:update-w-1}\\
\bm w_{in} &\gets \dfrac{\bm w_{in}}{\sqrt{ \bm w_{in}\Ht \bm U_{in} \bm w_{in} }}
,
\label{eq:update-w-2}
\intertext{where}
\bm U_{in} &= \dfrac1J \sum_j \dfrac1{c_{ijn}} \bm x_{ij}\bm x_{ij}\Ht,
\label{eq:update-w-U}\\
c_{ijn} &= {
\dfrac{\nu}{\nu+2} {\sigma_{ijn}}^2 + \dfrac{2}{\nu+2} \abs{y_{ijn}}^2
},
\label{eq:update-w-cijn}
\end{align}
and $\bm e_n$ is an $N$-dimensional vector
whose $n$th element is one and whose other elements are zero.
After calculating \eqref{eq:update-w-1} and \eqref{eq:update-w-2},
we update the separated signal by $y_{ijn}\gets \bm w_{in}\Ht \bm x_{ij}$.
In particular, when $\nu\to\infty$,
the majorization function \eqref{eq:w-update-majorization} converges to
the original cost function \eqref{eq:is-idlma-cost-function}, and
\eqref{eq:update-w-U} converges to
\begin{align}
\bm U_{in} &= \dfrac1J \sum_j \dfrac1{{\sigma_{ijn}}^2} \bm x_{ij}\bm x_{ij}\Ht.
\label{eq:update-w-U-is}
\end{align}
The update rule \eqref{eq:update-w-1}--\eqref{eq:update-w-cijn}
is equal to that in $t$-ILRMA.

To fix the scales of $y_{ijn}$ among the frequency bins,
the following back-projection technique is applied before updating $\bm \Sigma_n$ by \eqref{eq:update-dnn-1} and \eqref{eq:update-dnn-2}:
\begin{align}
 y_{ijn} \gets [\bm W_i^{-1}(\bm e_n\circ \bm y_{ij})]_{m_{\mathrm{ref}}},
\end{align}
where $ y_{ijn}$ is an element of ${\bm Y}_n$, $\circ$ is the Hadamard product,
$[\cdot]_n$ is the $n$th value of the vector,
and $m_{\mathrm{ref}}$ is the index of the reference channel.

\subsection{Relation between Parameter $\nu$ and Numerical Stability\label{sse:relation-stability}}
In Gauss-IDLMA, $\bm U_{in}$ defined by \eqref{eq:update-w-U-is} can be interpreted as
the spatial covariance matrix $\bm x_{ij} \bm x_{ij}\Ht$ weighted by ${\sigma_{ijn}}^{-2}$.
In general, $\sigma_{ijn}$ is estimated by $\mathrm{DNN}_n$,
whose output likely fluctuates, resulting in many spectral chasms in the time-frequency plane.
Therefore, the weight coefficient ${\sigma_{ijn}}^{-2}$ may be an excessively large value,
reducing the numerical stability of Gauss-IDLMA in IP.
In $t$-IDLMA, on the other hand,
$c_{ijn}$ in \eqref{eq:update-w-U} is the point internally dividing ${\sigma_{ijn}}^2$ and $\abs{y_{ijn}}^2$ with a ratio of $\nu:2$.
Since $y_{ijn}$ is the output of a linear filter,
$\abs{y_{ijn}}^2$ contains fewer chasms than ${\sigma_{ijn}}^2$;
this yields a beneficial spectral smoothing and numerical stability in optimization.

A prospective drawback of $t$-IDLMA is slower convergence,
especially in the case of small $\nu$ close to unity,
because the strong inference of DNN is discounted.
Thus, there is a tradeoff when setting $\nu$.
The appropriate selection of $\nu$ will be discussed in the next section.


\section{Experimental Evaluation}

\subsection{Task, Dataset, and Conditions}
We confirmed the validity of the proposed method
by conducting a music source separation task.
We compared four methods:
ILRMA (blind, $K=20$), DNN+WF, Duong+DNN,
and proposed IDLMA,
where DNN+WF applies a Wiener filter constructed
using all the outputs of the DNN source models
to the observed monaural signal~\cite{Uhlich2015_SingleSSwDNN}.
Note that MNMF was not included in this experiment
because its performance is almost always inferior to
that of ILRMA\cite{Mogami2017_tILRMA_ASJ}.
For Duong+DNN and IDLMA, the variance matrix $\bm\Sigma_n$ was updated by $\mathrm{DNN}_n$
after every 10 iterations of the spatial optimization.

We used the DSD100 dataset of SiSEC2016~\cite{Antoine2017_SiSEC2016}
as the dry sources and the training datasets of DNN,
where only bass (Ba.), drums (Dr.), and vocals (Vo.) were used in this experiment.
The 50 songs in the \texttt{dev} data were used to train $\mathrm{DNN}_n$
and the top 25 songs in alphabetical order in the \texttt{test} data were used
for performance evaluation.
The test songs were trimmed only in the interval of 30 to $\SI{60}{s}$.
To simulate a reverberant mixture,
we produced the two-channel observed signals
by convoluting the impulse response E2A ($T_{60}=\SI{300}{ms}$)
obtained from the RWCP database~\cite{SNakamura2000_RWCP} with each source,
and the mixture of Ba. and Vo. (Ba./Vo.) or Dr. and Vo. (Dr./Vo.) was separated.
The recording condition of E2A is given in~\cite{Mogami2017_tILRMA}.
All the signals were downsampled to $\SI{8}{kHz}$.
STFT was performed using a 512-ms-long Hamming window with a 256-ms-long shift
in the Ba./Vo. case
and a 256-ms-long Hamming window with a 128-ms-long shift in the Dr./Vo. case.
We used the signal-to-distortion ratio (SDR)~\cite{EVincent2006_BSSEval} as the total separation performance.

%

\subsection{Architecture and Training of DNN Source Model\label{sse:dnn-spectral-models}}
We constructed a fully connected DNN with four hidden layers.
Each layer had 1024 units, and
a rectified linear unit was used for the output of each layer.
To prepare the training data of mixture signals, we defined the following vectors:
\begin{align}
\vec{\bm s}_{jn}
&= (
\tilde{\bm s}_{(j-2c)n}\T,
\tilde{\bm s}_{(j-2c+2)n}\T,
\cdots,
\tilde{\bm s}_{(j+2c)n}\T
)\T\!
\in\cset^{I(2c+1)},\\
\vec{\bm x}_j
&= \p{
  \textstyle\sum_n \alpha_{jn} \vec{\bm s}_{jn}
  }\p{
  \norm{\textstyle\sum_n \alpha_{jn} \vec{\bm s}_{jn}}_2 + \delta_2
}^{-1}
\in \cset^{I(2c+1)}, \\
\bar{\bm s}_{jn}
&= \p{
  \alpha_{jn} \tilde{\bm s}_{jn}
  }\p{
  \norm{\textstyle\sum_n \alpha_{jn} \vec{\bm s}_{jn}}_2 + \delta_2
}^{-1}
\in \cset^{I},
\end{align}
where
$\tilde{\bm s}_{jn}\in\cset^I$ is the STFT of the $n$th source at $j$ (the column vector of $\tilde{\bm S}_n$),
$\vec{\bm x}_j$ and $\bar{\bm s}_{jn}$ are the mixture and source vectors, respectively,
$\alpha_{jn}$ is a random variable in the range $[0.05, 1]$,
which controls the signal-to-noise ratio in $\vec{\bm x}_j$,
and $\delta_2$ is a small value to avoid division by zero.
The input and output vectors of $\mathrm{DNN}_n$ are
$\abs*{\vec{\bm x}_j}^{.1}$ and $\abs*{\bar{\bm s}_{jn}}^{.1}$, respectively.

To optimize DNN, we added the term $\p{\lambda/2}\sum_q {g_q}^2$ to
\eqref{eq:is-idlma-dnn-loss} or \eqref{eq:t-idlma-dnn-loss}
for regularization, where $g_q$ is the weight coefficient in DNN,
and ADADELTA~\cite{Zeiler2012_ADADELTA} with a 128-size mini-batch was performed for 200 epochs.
The parameter $ \3$ was experimentally optimized
and set to $0.1\times(IJ)^{-1}\sum_{i,j}\hat r_{ijn}$.
The other parameters were set to
$ \delta_1=\delta_2=10^{-5} $,
$ c=3 $,
and
$
\lambda=10^{-5}
$.


\begin{figure}[tp]
\centering
\includegraphics[width=\columnwidth]{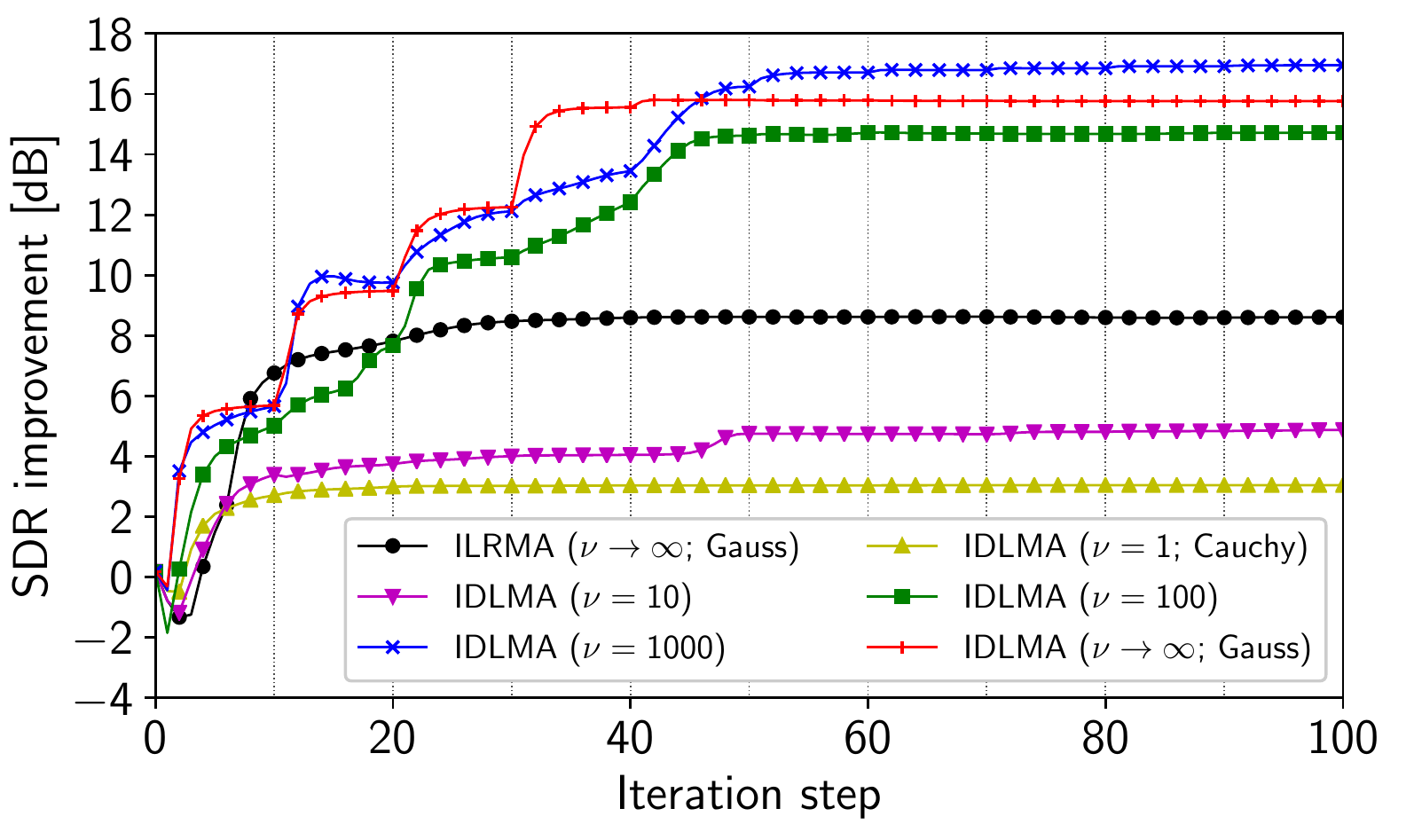}
\caption{Example of SDR improvements for each method for Ba./Vo.
\label{fig:example-sdr-improvement}%
}
\vspace{-1ex}
\end{figure}

\begin{figure}[tp]
\centering
\includegraphics[width=\columnwidth]{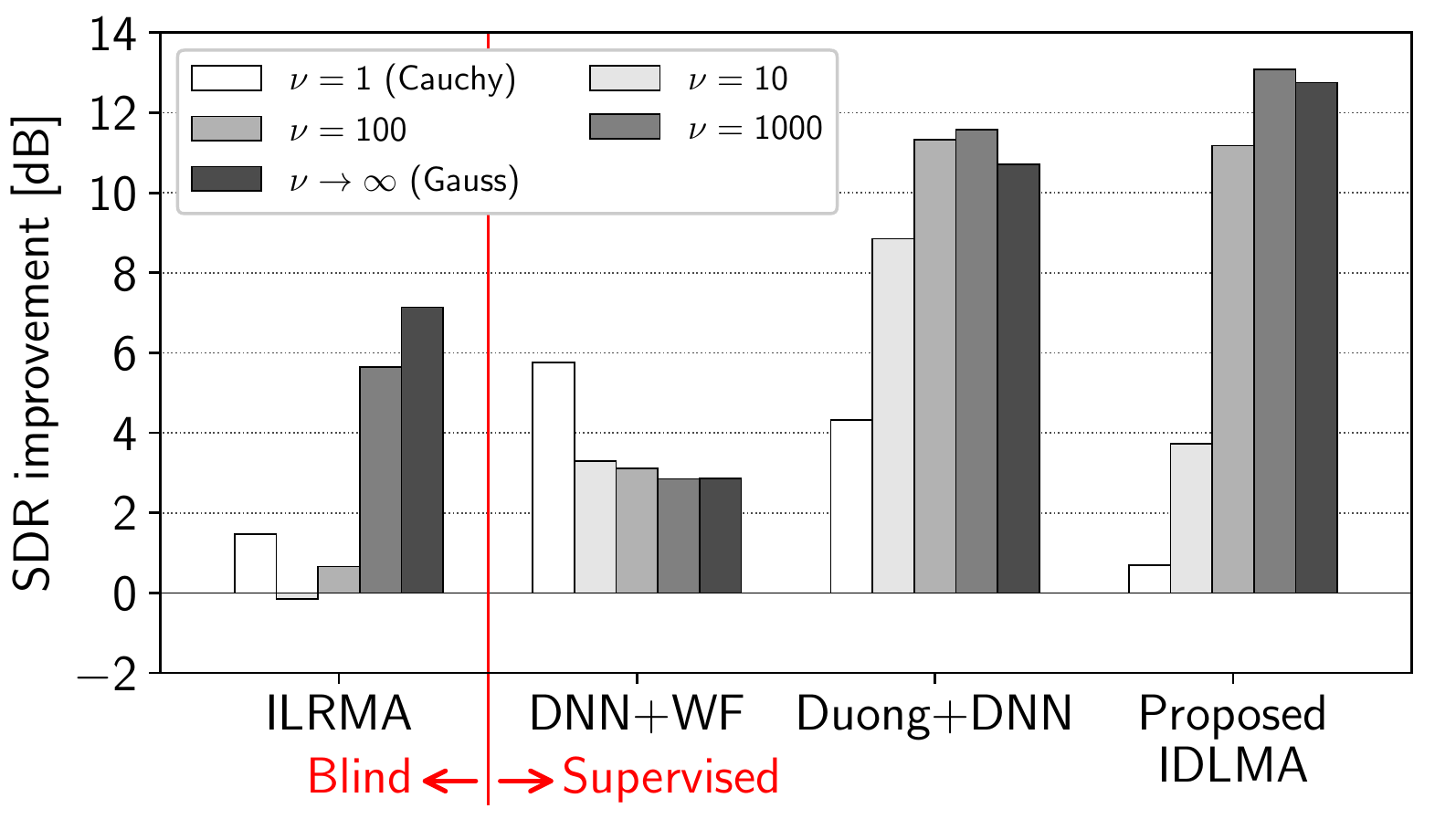}
\caption{Average SDR improvements of 25 Ba./Vo. songs.
\label{fig:result-vb}%
}
\vspace{-1ex}
\end{figure}

\begin{figure}[tp]
\centering
\includegraphics[width=\columnwidth]{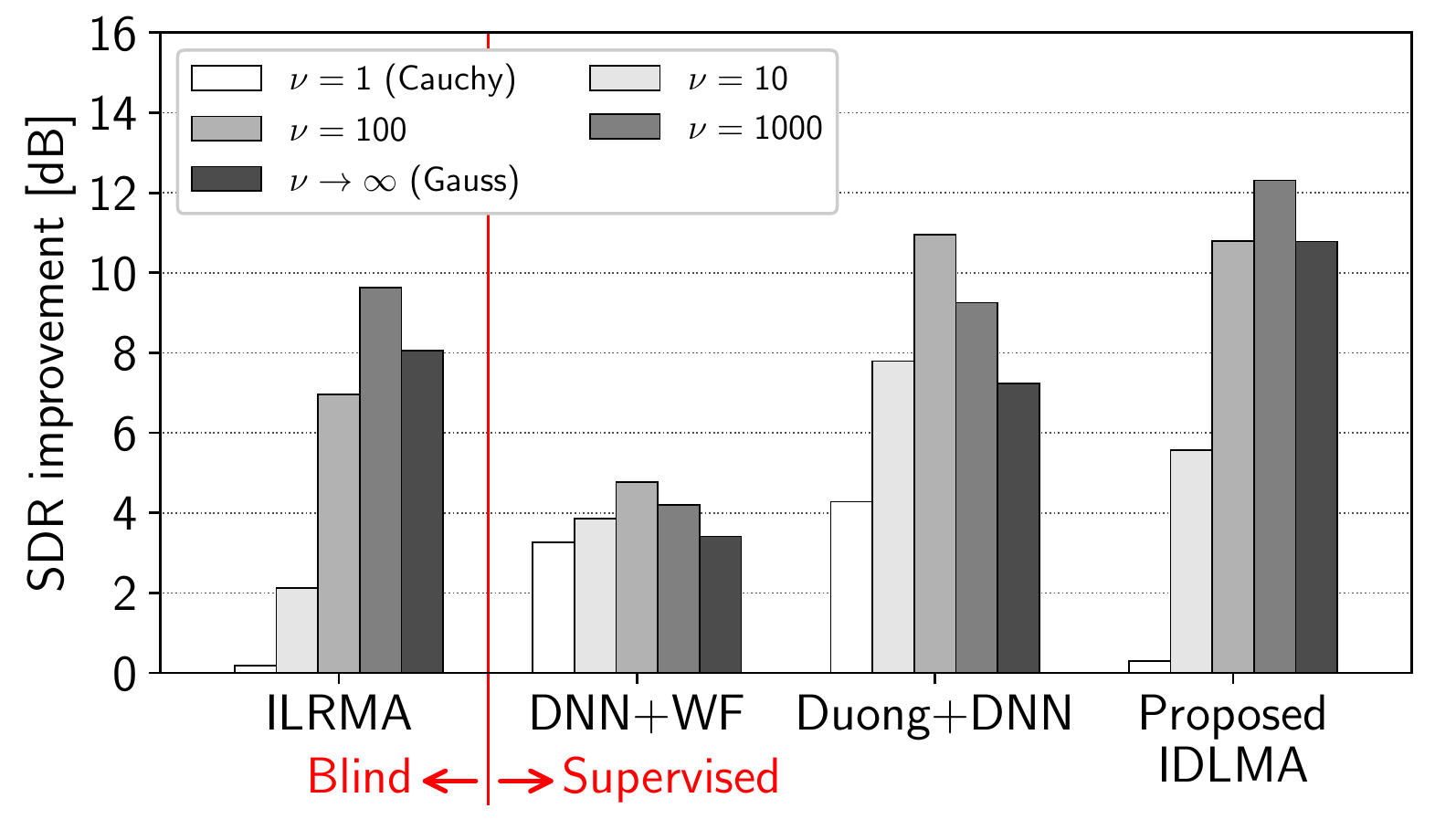}
\caption{Average SDR improvements of 25 Dr./Vo. songs.
\label{fig:result-vd}%
}
\vspace{-1ex}
\end{figure}

\subsection{Comparison of Separation Performance}
Fig.~\ref{fig:example-sdr-improvement} depicts an example of the convergence behaviors of ILRMA and IDLMA.
These results show that
(a) the DNN source model leads the demixing matrix to more accurate estimation,
resulting in a significant leap of SDR improvement, and
(b)
a larger $\nu$ provides a faster spatial model update
but $t$-IDLMA with the appropriate $\nu$ ($\mathord{=}1000$) converges to a higher SDR
than Gauss-IDLMA ($\nu=\infty$),
as mentioned in Sect.~\ref{sse:relation-stability}.

Figs.~\ref{fig:result-vb} and~\ref{fig:result-vd}
show the average SDR improvements of 25 test songs for Ba./Vo. and Dr./Vo.,
respectively.
We can confirm that
the proposed IDLMA outperforms the other methods for both mixtures of instruments.
In particular, $t$-IDLMA with $\nu=1000$ achieves the highest separation accuracy.

\subsection{Computational Times}
To show the efficiency of the proposed approach,
we compared the computational times of ILRMA, Duong+DNN, and IDLMA
for 100 iterations of spatial optimization.
We used Python 3.5.2 (64-bit) and Chainer 2.1.0 with an Intel Core i7-6850K ($\SI{3.60}{GHz}$, 6 Cores) CPU.
To calculate the DNN outputs, a GeForce GTX 1080Ti GPU was utilized.
Examples of computational times were
$\SI{23.3}{s}$ for ILRMA,
$\SI{287.1}{s}$ for Duong+DNN,
and $\SI{26.6}{s}$ for IDLMA.
These results confirm that the proposed method is as fast as conventional ILRMA
and more than 10 times faster than Duong+DNN.


\section{Conclusion}
In this paper, we proposed a new determined source separation method
that unifies ICA-based blind spatial optimization
and the DNN-based supervised source spectrogram model.
The proposed method employs a complex Student's $t$-distribution as the source generative model.
An experimental comparison showed
the efficacy of the proposed method in terms of
both the separation accuracy and the computational cost.


\section*{Acknowledgment}
This work was partly supported by
SECOM Science and Technology Foundation
and JSPS KAKENHI Grant Numbers
JP16H01735, 
JP17H06101, 
and JP17H06572. 

\end{document}